\documentclass[%
 aip,
 amsmath,amssymb,
 reprint,%
author-year,%
]{revtex4-1}

\usepackage{graphicx}
\usepackage{dcolumn}
\usepackage{bm}

\usepackage[utf8]{inputenc}
\usepackage[T1]{fontenc}
\usepackage{mathptmx}
\usepackage{etoolbox}

\makeatletter
\def\@email#1#2{%
 \endgroup
 \patchcmd{\titleblock@produce}
  {\frontmatter@RRAPformat}
  {\frontmatter@RRAPformat{\produce@RRAP{*#1\href{mailto:#2}{#2}}}\frontmatter@RRAPformat}
  {}{}
}%
\makeatother

\usepackage{hyperref}

\usepackage{color}
\usepackage{soul}

\newcommand{\azur}[2]{\st{#1} {\color{red}{#2}}}
\newcommand{\clara}[2]{\st{#1}{\color{green}{#2}}}

\begin{document}

\preprint{AIP/123-QED}

\title[Spectral response between particle and fluid kinetic energy in decaying homogeneous isotropic turbulence]{Spectral response between particle and fluid kinetic energy in decaying homogeneous isotropic turbulence}

\author{M. Schi\o{}dt}
    \email{maschi@dtu.dk}
\author{A. Hod\v zi\'c}%
\affiliation{ 
Technical University of Denmark, Kongens Lyngby, Denmark
}%

\author{F. Evrard}
\altaffiliation{Sibley School of Mechanical and Aerospace Engineering, Cornell University, Ithaca, United States}
\author{M. Hausmann}
\author{B. Van Wachem}
\affiliation{%
Otto von Guericke University, Magdeburg, Germany 
}%

\author{C.M. Velte}%
\affiliation{ 
Technical University of Denmark, Kongens Lyngby, Denmark
}%

\date{\today}

\begin{abstract}
In particle-laden turbulence, the Fourier Lagrangian spectrum of each phase is regularly computed, and analytically derived response functions relate the Lagrangian spectrum of the fluid- and the particle phase. However, due to the periodic nature of the Fourier basis, the analysis is restricted to statistically stationary flows. In the present work, utilizing the bases of time-focalized proper orthogonal decomposition (POD), this analysis is extended to temporally non-stationary turbulence. Studying two-way coupled particle-laden decaying homogeneous isotropic turbulence for various Stokes numbers, it is demonstrated that the temporal POD modes extracted from the dispersed phase may be used for the expansion of both fluid- and particle velocities. The POD Lagrangian spectrum of each phase may thus be computed from the same set of modal building blocks, allowing the evaluation of response functions in a POD frame of reference. Based on empirical evaluations, a model for response functions in non-stationary flows is proposed. The related energies of the two phases is well approximated by simple analytical expressions dependent on the particle Stokes number. It is found that the analytical expressions closely resemble those derived through Fourier analysis of statistically stationary flows. These results suggest the existence of an inherent spectral symmetry underlying the dynamical systems consisting of particle-laden turbulence, a symmetry which spans across stationary/non-stationary particle-laden flow states. 
\end{abstract}

\maketitle

\section{Introduction} \label{sec:Introduction}
Recent years have seen renewed attention directed towards particle-laden turbulence, due to its relevance in numerous engineering and natural settings (\cite{brandt2022review}). Theoretical models and improved experimental and numerical methods have led to advancements in our understanding of particle dynamics, herein counting acceleration statistics, preferential sampling and particle clustering to name a few (\cite{toschi2009review}; \cite{gustavsson2016review}; \cite{maxey2017review}).

One focus of study has been the modulation of turbulence induced by two-way coupling (\cite{Druzhinin1999}; \cite{Ferrante2003}). Here, the presence of particles in flows under zero gravity conditions has been shown to attenuate turbulent kinetic energy (TKE) at low wavenumbers and augment it at higher wavenumbers, leading to an increase in dissipation (\cite{squires1994}). Inertial particles may, however, also act as sources of increased turbulence energy, and the total TKE may be either augmented or attenuated by the presence of a dispersed phase (\cite{Ferrante2003}). A key parameter identified in this regard is the particle Stokes number. \cite{Letournel2020} investigated TKE totals as a function of the Stokes number, and found an approximate threshold below which turbulence was augmented, and above which it was attenuated. Nevertheless, the same authors underlined the lack of consensus on a unique criterion for turbulence modulation by particles.

\cite{ireland2016part1} investigated the large scale single-particle velocity statistics of inertial particles in homogeneous isotropic turbulence (HIT). Driven by the effects of inertial filtering and preferential sampling, the average particle kinetic energy normalized by the average fluid kinetic energy was shown to approximately follow a simple relation dependent on the Stokes number. Similar studies were conducted under gravity conditions by \cite{good2014} and \cite{ireland2016part2}. 

Although the study of particle-laden turbulence has rapidly progressed over the past decade, new theoretical tools are still needed in order to gain further insights into the dynamics (\cite{brandt2022review}). One such tool may be the particle proper orthogonal decomposition (PPOD) formulated by \cite{schiodt2022}, where Lagrangian particle velocities are decomposed into a set of modes that represent temporal particle dynamics. This tool is utilized in the present study, where the extracted modes are compared to those extracted for the fluid measured at fixed Eulerian mesh points using the temporal formulation of POD introduced by \cite{aubry1991spatiotemporal}. Both formulations of POD are briefly outlined in \autoref{sec:POD}, and the constraints required for direct comparisons of fluid- and particle POD modes are listed. 

Modal decomposition of fluid- and particle temporal dynamics allows for the evaluation of the Lagrangian spectrum of both phases in a POD frame of reference. In the current work, this leads to formulations of POD-based response functions, that relate the energy of the two phases on a modal level. Although response functions based on the Fourier decomposition have previously been studied in stationary flows (\cite{csanady1963}; \cite{zhang2019}; \cite{berk2021}), the advantage of the POD-based approach is that stationarity is not required, and the present study is therefore focused on the analysis of various simulations of two-way coupled particle-laden decaying HIT. The analysis culminates in analytic expressions of POD-based response functions closely resembling those derived through Fourier analysis of stationary flows. 

Section \ref{sec:POD} gives a brief outline of the formulation of POD and the structure of the ensembles that will produce temporal modes representing fluid- and particle dynamics. A summary of the simulation setup is given in \ref{sec:Simulation}, which is followed by a presentation and discussion of results in section \ref{sec:Results}. Finally, our conclusions are given in section \ref{sec:Conclusion}.

\section{Proper orthogonal decomposition} \label{sec:POD}
The main objective of POD is to extract a set of empirical basis functions $\boldsymbol{\varphi} = \{ \varphi _\alpha \}_{\alpha=1}^M$ that represent dominating features of the studied dynamical system. The basis functions, also known as modes, are extracted by solving the eigenvalue problem
\begin{align}
    \mathcal{R} \varphi _\alpha = \lambda _\alpha \varphi _\alpha, \quad \alpha \in [1:M], \label{eq:eigenproblem}
\end{align}
where $\boldsymbol{\lambda} = \{ \lambda _\alpha \}_{\alpha=1}^M$ are the eigenvalues connected to each mode, and for the cases we study, these are real and sorted such that $\lambda _1 \geq \cdots \geq \lambda_M \geq 0$. The operator $\mathcal{R}: \mathcal{H} \rightarrow \mathcal{H}$ is defined from the ensemble of empirical data  $\boldsymbol{u} = \{ u^{(i)} \}_{i=1}^{N_e}$, and is dependent on the definition of the Hilbert space $\mathcal{H}$ for which $\boldsymbol{\varphi}$ serves as an empirical orthonormal basis. Though the basis is not necessarily complete in $\mathcal{H}$, each ensemble member may be decomposed into a weighted sum of modes, thus
\begin{align}
    u^{(i)} = \sum \limits_{\alpha=1}^M c_\alpha ^{(i)} \varphi _\alpha, \quad i \in [1:N_e],
\end{align}
where the weights $c_\alpha ^{(i)}$ are known as the projection coefficients given by
\begin{align}
    c_\alpha ^{(i)} = ( u^{(i)}, \varphi _\alpha ). \label{eq:Coef}
\end{align}
Here $(\cdot, \cdot)$ denotes the inner product of $\mathcal{H}$. The projection coefficients are connected to the eigenvalues $\boldsymbol{\lambda}$ by the relation
\begin{align}
    \lambda _\alpha = \left\langle \left\{ c^{(i)}_\alpha c^{(i)*}_\alpha  \right\}_{i=1}^{N_e} \right\rangle, \quad \alpha \in [1:M], \label{eq:EigCoef}
\end{align}
where $(^*)$ denotes both the complex conjugate transpose for a scalar and Hermitian transpose for a vector, and $\langle \{ \cdot \}_{i=1}^{N_e} \rangle$ is the ensemble average operator.

The definition of $\mathcal{H}$ and what constitutes an ensemble member determines the interpretation of $\boldsymbol{\varphi}$ and $\boldsymbol{\lambda}$. In \autoref{sec:eulerPOD} and \autoref{sec:PPOD} we briefly outline the discrete formulations of the Eulerian- and the Lagrangian (particle) POD, respectively, and show their dependency on the definition of $\boldsymbol{u}$.

\subsection{Eulerian POD} \label{sec:eulerPOD}
The most common application of POD is based on the fluid velocity $\boldsymbol{u}_f(\boldsymbol{x}, t) \in \mathbb{R}^D$ measured at fixed mesh points in a Eulerian grid at equidistant sample times. Following the classical interpretation of POD (\cite{lumley1967}) an ensemble member may in this discrete case be formed by 
\begin{align}
    \boldsymbol{u}^{(i)} = \begin{bmatrix} 
    \boldsymbol{u}_f^{(i)}(\boldsymbol{x}_{1}, t_0)^* & \cdots & \boldsymbol{u}_f^{(i)}(\boldsymbol{x}_{N_g}, t_0)^* & \cdots & \boldsymbol{u}_f^{(i)}(\boldsymbol{x}_{N_g}, t_{N_t-1})^*
    \end{bmatrix}^*.
\end{align}
Here $\boldsymbol{u}_f ^{(i)}$ is the $i$'th fluid velocity realization, $\boldsymbol{x}_g \in \mathbb{R}^D$, $g \in [1:N_g]$ are the Eulerian mesh points and $t_n \in T$, $n \in [0:N_t-1]$ are the sample times of the temporal domain $T$. In this case $\boldsymbol{u}^{(i)} \in \mathcal{H} = \mathbb{R}^N$, where $N=D N_g N_t$. $\mathcal{H}$ is equipped with the standard inner product $(\boldsymbol{w}_1, \boldsymbol{w}_2) = \boldsymbol{w}_ 2^* \boldsymbol{w}_1$, and the operator $\mathcal{R}$ in equation \eqref{eq:eigenproblem} is given by
\begin{align}
    \mathcal{R} = \left\langle \left\{ \boldsymbol{u}^{(i)} \boldsymbol{u}^{(i)*}  \right\}_{i=1}^{N_e} \right\rangle \in \mathbb{R}^{N \times N}. \label{eq:Roperator}
\end{align}
Solving equation \eqref{eq:eigenproblem} then results in a set of spatio-temporal modes that are optimal with respect to energy, where $\boldsymbol{\lambda}$ represents the energy of each mode. However, the amount of data needed to generate $\boldsymbol{\varphi}$ often makes this classical approach infeasible, as several uncorrelated fluid flow realizations are needed to generate the data. Instead, an approach popularized by \cite{sirovich1987} and \cite{aubry1988} is to extract spatially orthogonal modes, with time dependent projection coefficients. This is what \cite{towne2018} refers to as the \textit{space-only} POD, and in a statistically stationary flow an ensemble member may be given by the fluid velocity measured at all grid points at a single sample time. From one fluid realization several ensemble members may thus be generated, and the ensemble average operator reduces to a temporal average. 

In the current work we will focus on what we term the \textit{time-only} POD (TPOD) and its relation to PPOD. The TPOD is also formulated in the continuous case (\cite{aubry1991spatiotemporal}; \cite{aubry1991beauty}) and as an analogy to its spatial counterpart it produces a set of temporally orthogonal modes, with spatially dependent projection coefficients. An ensemble member is in this case given by
\begin{align}
    \boldsymbol{u}^{(i)} = \begin{bmatrix} 
    \boldsymbol{u}_f(\boldsymbol{x}_{i}, t_0)^* & \cdots & \boldsymbol{u}_f(\boldsymbol{x}_{i}, t_{N_t-1})^*
    \end{bmatrix}^*, \quad i \in [1:N_e],
\end{align}
i.e. the fluid velocity at a grid point $i$ measured at sample times $t_n$. Note that $N_e \leq N_g$ when the ensemble members are taken from the same fluid realization, and that the fluid flow in that case should be homogeneous (\cite{aubry1991beauty}), signifying that the temporal evolution is statistically equivalent in all grid points. The ensemble average operator then reduces to a spatial average and the operator $\mathcal{R}$ is still given as in equation \eqref{eq:Roperator}, although here $N = D N_t$.

The modes extracted with TPOD represent the temporal evolution of the fluid velocity through a Eulerian mesh point, and $\boldsymbol{\lambda}$ is connected to the energy
\begin{align}
    E(t) = \frac{1}{2} \left\langle \left\{ \boldsymbol{u}_f ^*(\boldsymbol{x}_i, t) \boldsymbol{u}_f (\boldsymbol{x}_i, t) \right\}_{i=1}^{N_e} \right\rangle \label{eq:EEnergy}\,,
\end{align}
by 
\begin{align}
    \sum \limits _{n=0}^{N_t-1} E(t_n) = \frac{1}{2} \sum \limits_{\alpha = 1}^M \lambda _\alpha\,. \label{eq:Energy}
\end{align}

\subsection{Particle POD} \label{sec:PPOD}
\cite{schiodt2022} formulated PPOD as a method for decomposing the velocity of Lagrangian particles into a weighted sum of empirical modes. Like TPOD the method produces a set of temporal modes, however, the modes represent the dynamics of Lagrangian particles rather than the fluid dynamics at fixed Eulerian mesh points. The ensemble $\boldsymbol{u}$ is in this formulation defined by the ensemble members
\begin{align}
    \boldsymbol{u}^{(i)} = \begin{bmatrix} \boldsymbol{v}^{(i)} (t_0) ^* & \cdots & \boldsymbol{v}^{(i)} (t_{N_t-1}) ^* \end{bmatrix}^*, \quad i \in [1:N_e]\,,
\end{align}
where
\begin{align}
    \boldsymbol{v}^{(i)}(t_n) = \begin{bmatrix} \boldsymbol{v}^{(i)}_1(t_n)^* & \cdots & \boldsymbol{v}^{(i)}_{N_p}(t_n)^* \end{bmatrix} ^*\,,
\end{align}
is the velocity of $N_p$ Lagrangian particles measured at sample times $t_n \in T$, $n \in [0:N_t-1]$. Here $\boldsymbol{u}^{(i)} \in \mathbb{R}^N$ with $N = D N_p N_t$, since $\boldsymbol{v}_p ^{(i)}(t) \in \mathbb{R}^D$ is the velocity of a single particle. Choosing $N_p = 1$ for the remainder of the current work, we see that PPOD and TPOD ensemble members belong to the same Hilbert space $\mathcal{H}=\mathbb{R}^N$, $N=DN_t$. The mode-sets extracted with respectively TPOD and PPOD are therefore in this case directly comparable.

To generate a meaningful ensemble of Lagrangian particle velocities, the ensemble particles should belong to similar flows or be sampled from the same flow containing certain symmetries. We elaborate further on this point in \autoref{sec:d-HIT}.

In \autoref{sec:Results} both TPOD and PPOD analysis is applied to the Reynolds decomposed $\boldsymbol{u}_{fluct}^{(i)} = \boldsymbol{u}^{(i)} - \langle \{ \boldsymbol{u}^{(i)} \} _{i=1}^{N_e} \rangle$ rather than $\boldsymbol{u}^{(i)}$. Thus, $E(t)$ in equations \eqref{eq:EEnergy}-\eqref{eq:Energy} becomes a measure of TKE, and 
\begin{align}
    \boldsymbol{u}^{(i)} = \left\langle \left\{ \boldsymbol{u}^{(i)} \right\}_{i=1}^{N_e} \right\rangle + \sum \limits_{\alpha=1}^{M} c^{(i)} _{\alpha} \boldsymbol{\varphi} _\alpha, \quad i \in [1:N_e]\,.
\end{align}
However, $\langle \{ \boldsymbol{u}^{(i)} \}_{i=1}^{N_e} \rangle \approx \boldsymbol{0}$ for all TPOD and PPOD ensembles considered, and we will therefore interchangeably refer to $\boldsymbol{\varphi}$ as the mode-set spanning both the signal $\boldsymbol{u}^{(i)}$ and $\boldsymbol{u}_{fluct}^{(i)}$. 

\section{Simulation} \label{sec:Simulation}
In the current work we consider the simulation of one single-phase flow and four different simulations of two-way coupled particle-laden turbulence. All simulations are performed within a periodic cube with edge length $\ell$, discretized into $N_g$ computational cells. 

\subsection{Dynamical equations}
We apply the Euler-Lagrange point-particle approach (\cite{Elghobashi1992}) where the fluid velocity $\boldsymbol{u}_f$ is computed at each time step by numerical integration of the incompressible Navier-Stokes equations on a Eulerian mesh, and particle velocities are obtained by integrating the governing particle equations of motion forward in time. For the Navier-Stokes equations a constant dynamic viscosity $\mu _f$ and mass density $\rho _f$ are used, and with $\boldsymbol{p}$ denoting pressure the equations are given by
\begin{subequations}
\begin{align}
    \nabla \cdot \boldsymbol{u}_f &= 0\,, \\
    \frac{\partial \boldsymbol{u}_f}{\partial t} + \nabla \cdot (\boldsymbol{u}_f \otimes \boldsymbol{u}_f) &= - \frac{1}{\rho _f} \nabla \boldsymbol{p} + \frac{\mu _f}{\rho_f} \nabla ^2 \boldsymbol{u}_f + \boldsymbol{F}_p + \boldsymbol{F}\,.
\end{align}
\label{eq:FluidMotion}
\end{subequations}
Here $\boldsymbol{F}_p$ is the force that the dispersed particles exert on the carrier fluid, and $\boldsymbol{F}$ is an artificial source term applied in an initial forcing period. In \autoref{sec:Forcing} the details of $\boldsymbol{F}_p$ and $\boldsymbol{F}$ are outlined.

The particles considered are monodisperse solid spheres with diameter $d_p$, volume $V_p$ and density $\rho _p$. Assuming particles are only accelerated according to drag force, the dynamic equations for particle motion are given by
\begin{subequations}
\begin{align}
    \frac{\mathrm{d}\boldsymbol{x}_p}{\mathrm{d}t} &= \boldsymbol{v}_p\,, \\
    V_p \rho_p \frac{\mathrm{d}\boldsymbol{v}_p}{\mathrm{d}t} &= \boldsymbol{F}_D = \frac{\pi}{8} d_p^2 \rho_f C_D |\boldsymbol{u}_{f@p} - \boldsymbol{v}_p | ( \boldsymbol{u}_{f@p} - \boldsymbol{v}_p )\,,
\end{align}
\label{eq:ParticleMotion}
\end{subequations}
where $\boldsymbol{x}_p$ and $\boldsymbol{u}_{f@p}=\boldsymbol{u}_f(\boldsymbol{x_p}, t)$ are the particle position and fluid velocity at particle position, respectively. $\boldsymbol{F}_D$ denotes the drag force and $C_D$ is the drag coefficient given by (\cite{Schiller1933}) 
\begin{align}
    C_D = \frac{24}{Re_p} \left( 1 + 0.15 Re_p ^{0.687} \right)\,, \label{eq:DragCoeff}
\end{align}
and 
\begin{align}
    Re_p = \frac{d_p \rho _f | \boldsymbol{u}_{f@p} - \boldsymbol{v}_p |}{\mu _f}\,,
\end{align}
is the particle Reynolds number. Equation \eqref{eq:DragCoeff} holds for $0 < Re_p \leq 1000$, which is the only range considered in the current work. 

A second-order finite-volume solver (\cite{Denner2020}) is used to integrate \eqref{eq:FluidMotion} forward in time, and the Verlet scheme is used for the forward integration of \eqref{eq:ParticleMotion}.

\subsection{Decaying homogenous isotropic turbulence} \label{sec:d-HIT}
To study two-way coupling effects in an idealized test case, we analyze a particle-laden fluid with decaying HIT. This case is chosen over stationary HIT because the effects of the forcing term $\boldsymbol{F}$ would overlap with the particle-fluid interaction energy in the latter (\cite{Abdelsamie2012}). In addition, the properties of decaying HIT signifies that the fluid velocity in all Eulerian mesh points evolves in a statistically equivalent manner. The inertial particles are thermalized to the fluid (see \autoref{sec:Forcing}) and thus have a statistically equivalent evolution throughout the temporal domain. Therefore, a meaningful ensemble of realizations can be generated for both TPOD and PPOD from a single simulation of particle-laden turbulence. For TPOD, the ensemble members are formed by sampling the fluid velocity at $N_e$ equidistantly spaced mesh points at sample times $t_n \in T$, $n \in [0:N_t-1]$, and for PPOD the ensemble members are formed by randomly choosing $N_e$ particle records to track over the same sample times. The inertial particles are initially spaced randomly throughout the cubic domain in order to avoid introducing bias. 

\subsection{Forces} \label{sec:Forcing}
Each simulation can be split into two periods -- a forcing period, and a decaying period. The forcing period is the initial part of the simulation, in which HIT is obtained by applying the source term $\boldsymbol{F}$ in equation \eqref{eq:FluidMotion}. This period is necessary to initiate decay from a fully developed turbulent velocity field. The forcing procedure follows the forcing scheme developed by \cite{Mallouppas2013Force} and is the same as the one briefly outlined in \cite{schiodt2022}. 

During the forcing period particles are present within the fluid, but two-way coupling is deactivated, i.e. $\boldsymbol{F}_p = \boldsymbol{0}$ in equation \eqref{eq:FluidMotion}. This allows for the thermalization of particles under one-way coupling conditions, which minimizes the transitional regime when two-way coupling is activated (\cite{Ferrante2003}). 

We define the end of the forcing period as time $t_0=0$s, which also denotes the start of the decaying period. Here $\boldsymbol{F} = \boldsymbol{0}$, and two-way coupling is activated for the multiphase simulations, but remains zero for the single-phase simulation.

The two-way coupling term, $\boldsymbol{F}_p$, in equation \eqref{eq:FluidMotion} is modelled as suggested by \cite{crowe1977} where
\begin{align}
    \boldsymbol{F}_p = - \frac{1}{\rho _f V_{g}} \sum \limits_{p'=1}^{N_{p,g}} \boldsymbol{F}_{D,p'}\,.
\end{align}
Here $V_{g}$ is the volume of cell $g$ in the discretized domain, and $N_{p,g}$ is the number of particles present in that cell. $\boldsymbol{F}_{D,p'}$ is the drag force exerted by the fluid on particle $p'$. 

\subsection{Setup}
\subsubsection{Fluid}
We use the setup of \cite{Mallouppas2017dissipation} for the fluid simulation. Here the cube edge length is given by $\ell = 0.128$m, and the domain is discretized into $N_g = 128^3$ computational cells. Fluid viscosity is given by $\mu _f = 1.72 \times 10^{-5}$Pa s, and fluid density by $\rho _f = 1.17$kg m$^{-3}$. For all of the subsequent cases studied the Taylor Reynolds number at $t_0$ is given by $Re_\lambda = 58.0$, where the integral-, Taylor-, and Kolmogorov length scales are respectively $I=1.129 \times 10^{-2}$m, $\lambda = 6.134 \times 10^{-3}$m and $\eta = 4.0 \times 10 ^{-4}$m. The Kolmogorov time scale at $t_0$ is $\tau _\eta  = 10^{-2}$s. The reader is referred to \cite{schiodt2022} for a more thorough outline of the temporal evolution of the fluid characteristics in the single-phase simulation. 

\subsubsection{Particles}
The different multiphase simulations considered are characterized by the Stokes number $St (t)= \tau _p (t) / \tau _{\eta} (t)$ of the inertial particles at $t=t_0$. Here $\tau _p$ (equation \eqref{eq:ParticleResponseTime}) is the particle response time. The particle diameter is set to $d_p = 1.0 \times 10 ^{-4}$m, and the particle mass fraction $\phi _m \approx 1$. Since the particle density $\rho_p$ is tweaked in each case to obtain different Stokes numbers, this signifies that the number of particles present in the fluid varies between each case. Letting $St_0 = St(t_0)$, the Stokes numbers considered are $St_0 = 0.25$, $St_0 = 0.75$, $St_0 = 1.5$, and $St_0 = 3.0$.

\begin{figure}
    \centering
    \includegraphics[scale=0.5]{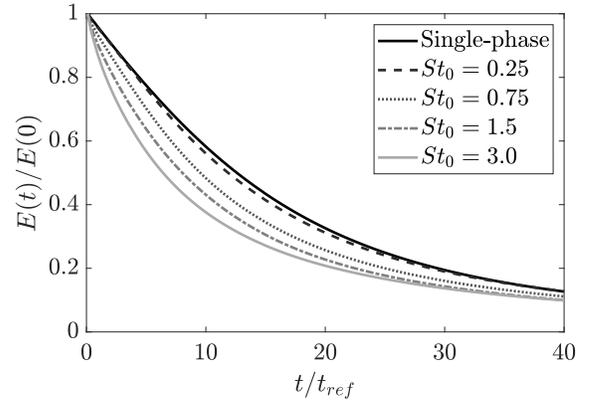}
    \caption{Evolution of normalized turbulent kinetic energy.}
    \label{fig:TKE}
\end{figure}

\section{Results \& discussion} \label{sec:Results}
All subsequent results are based on fluid- and inertial particle velocities during the decaying period, which lasts for $0.4\,\mathrm{s}$ of physical time. The velocities are sampled every $\delta _t = 10^{-3}$ seconds, amounting to $N_t = 400$ temporal samples. The temporal domain is normalized with respect to the reference time scale $t_{ref} = \tau _\eta (t_0) = 10^{-2}\,\mathrm{s}$ which is shared between all simulations. 

\subsection{Fluid statistics}
\autoref{fig:TKE} shows the temporal evolution of the carrier phase TKE, $E(t)$, in the single- and multiphase simulations. The TKE is normalized by $E(0)$, and the figure illustrates that turbulence is increasingly attenuated for increasing Stokes numbers. However, at $St_0 = 0.25$ there is a slight augmentation of turbulence for $t/t_{ref} > 38$. Similar observations have been reported in previous studies (\cite{sundaram1999}; \cite{Ferrante2003}; \cite{Letournel2020}).

\begin{figure}
    \centering
    \vspace{0.25cm}
    \includegraphics[scale=0.52]{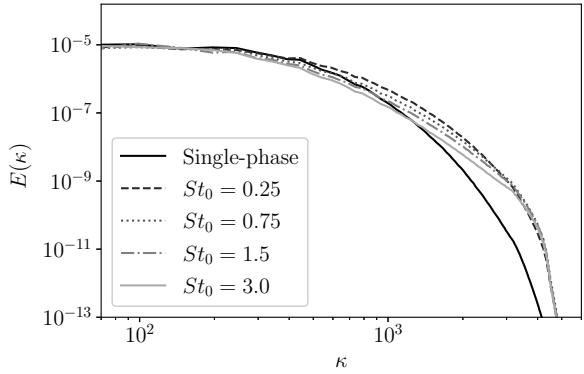}
    \caption{Fourier turbulence energy spectrum $E(\kappa)$ of each simulation at final time step $t/t_{ref} = 40$.}
    \label{fig:spectrum}
\end{figure}

The Fourier turbulence energy spectrum $E(\kappa)$ of the carrier phase at time $t/t_{ref} = 40$ is seen in \autoref{fig:spectrum}. As observed in previous work (\cite{Druzhinin1999}; \cite{Ferrante2003}; \cite{Letournel2020}) the presence of inertial particles modulates the spectrum, shifting energy from low to high wavenumbers. The degree with which this energy transfer occurs is dependent on the Stokes number, where more energy is observed to be transferred at lower Stokes numbers.

\begin{figure}
    \centering
    \includegraphics[scale=0.5]{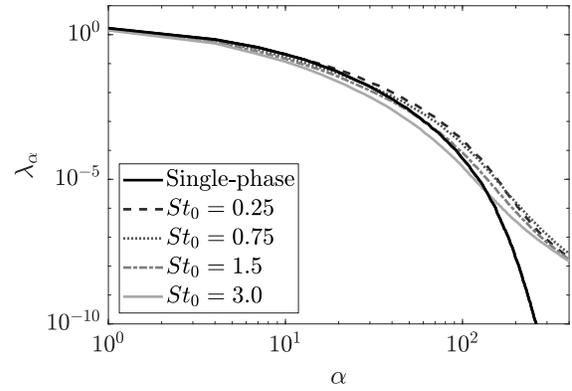}
    \caption{TPOD eigenspectrum showing the distribution of modal energy of the carrier phase in each simulation case.}
    \label{fig:TPODEnergy}
\end{figure}

Increased energy at high wavenumbers implies more energetic small scale turbulence structures. The fluid velocity measured over time at a fixed spatial point will therefore, on average, contain more fluctuations for the multiphase flows compared to the single-phase flow. This behaviour is indeed observed when considering the TPOD eigenspectra of \autoref{fig:TPODEnergy}. Here, the extracted modes $\boldsymbol{\varphi}$ and corresponding eigenvalues $\boldsymbol{\lambda}$ are based on the 3-D fluid velocity measured in $N_e = 16^3=4096$ equidistantly spaced Eulerian mesh points, where these ensemble members are assumed to represent the dynamics of all $128^3$ mesh points (see \autoref{sec:d-HIT}).

\autoref{fig:TPODEnergy} shows, for all simulations, the energy $\lambda _\alpha$ of each TPOD mode for $\alpha \in [1:400]$. A brief glance at the eigenspectra depicted reveals a distinct difference of shape between the single- and multiphase simulations. The figure also illustrates that modal energy is slightly higher in the single-phase case when the mode number is low, whereas for higher mode numbers the modal energy is higher in the multiphase cases. As will be shown later (\autoref{fig:Modes}) the higher numbered modes contain more fluctuations, and this observation therefore aligns well with the intuition of how TPOD modal energy should be distributed in accordance to the spatial structures. It is notable that the modal energy is larger for some mode numbers in the multiphase cases compared to the single-phase case even though the total modal energy in the latter is larger (see \autoref{fig:TKE}). This further underlines the observation that a larger fraction of energy is distributed to more rapidly fluctuating TPOD modes when the fluid is laden with inertial particles and two-way coupling is activated. 

\subsection{PPOD convergence}
PPOD is applied to the velocity of $N_e = 4096$ randomly selected inertial particles, initially distributed throughout the spatial domain. This is performed for all multiphase simulations under the assumption that these subsets of particles represent the dynamics of all particles within each respective simulation.
\begin{figure}[b]
    \centering
    \includegraphics[scale=0.5]{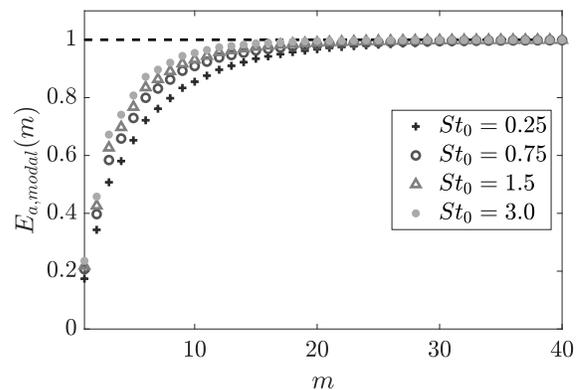}
    \caption{Convergence of PPOD accumulated modal energy.}
    \label{fig:ConvergenceStokes2}
\end{figure}
\begin{figure*}
    \centering
    \includegraphics[scale=0.50]{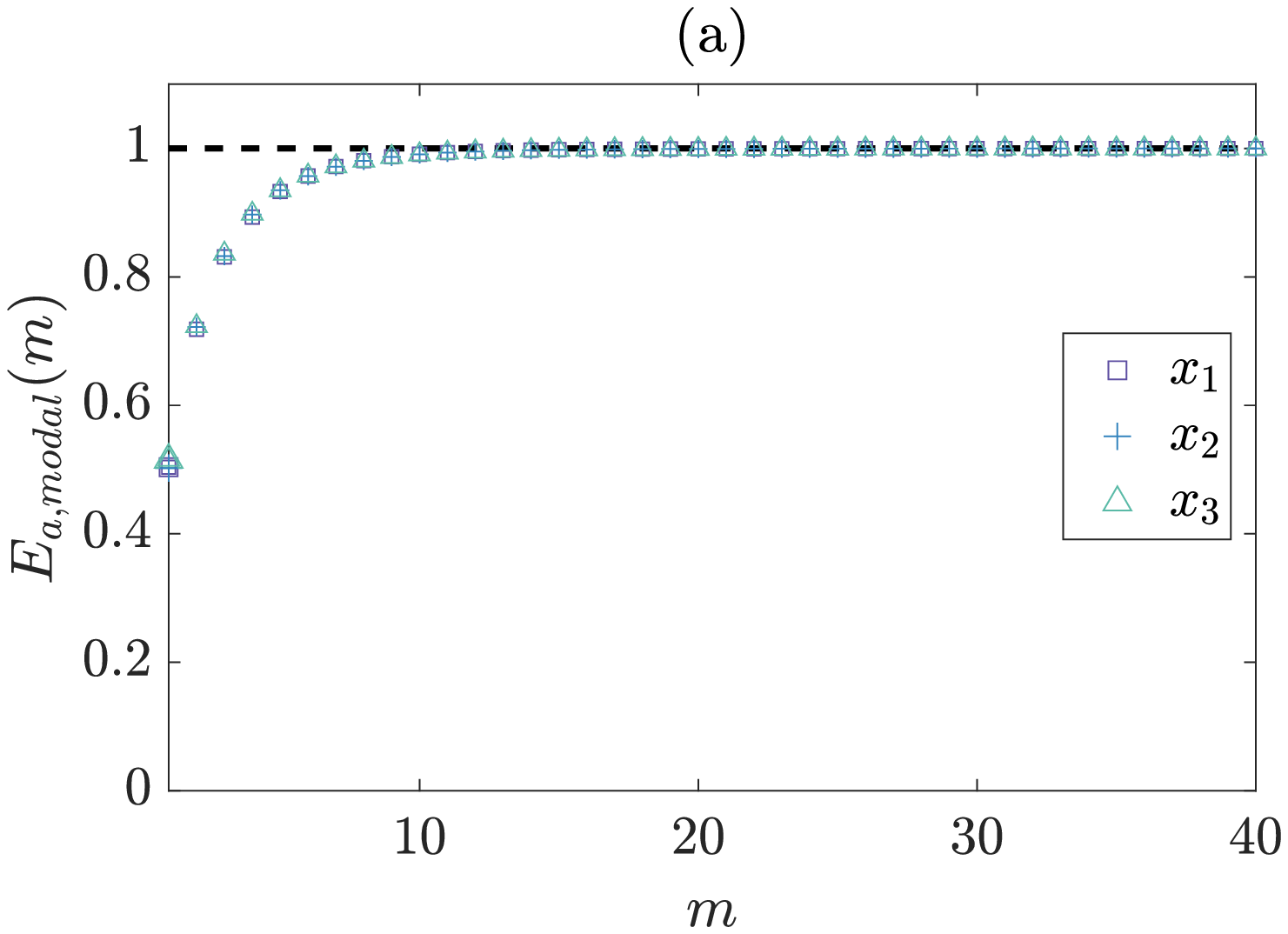}
    \includegraphics[scale=0.50]{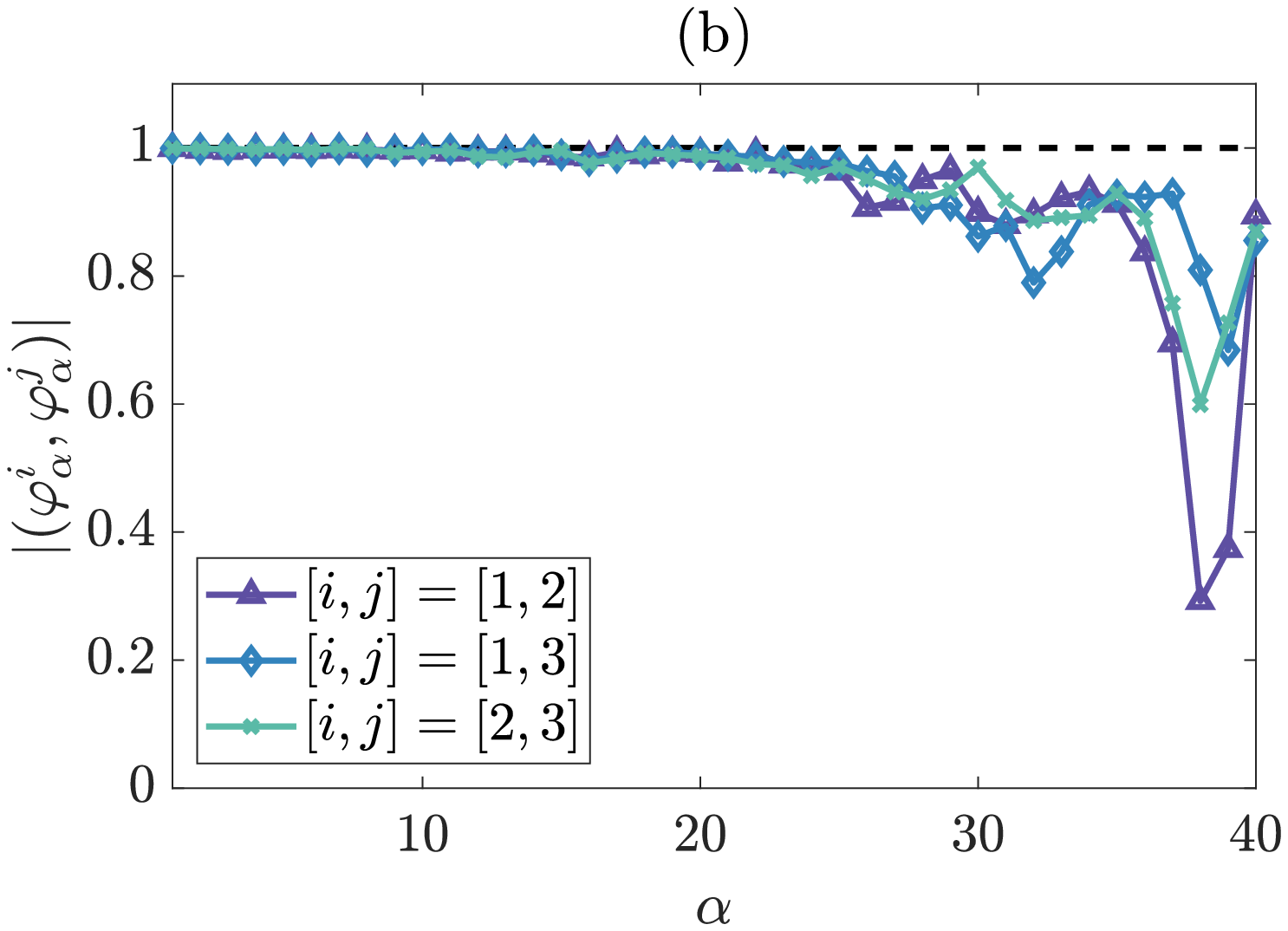}
    \caption{(a) Convergence rates of $E_{a,modal}$ are equivalent between each velocity component and (b) the extracted modes are almost completely parallel for $\alpha \leq 20$.}
    \label{fig:Component}
\end{figure*}
Let $E_{a,modal}(m)$ denote the fraction of accumulated POD modal energy up until mode number $m$:
\begin{align}
    E_{a,modal}(m) = \sum \limits_{\alpha=1}^{m} \lambda _\alpha \Bigg/ \sum \limits_{\beta=1}^{M} \lambda _\beta, \quad m \in [1:M].
\end{align}
Although $m \in [1:M]$, $M = \min (N, N_e) = 1200$, the statistic is only shown for $m \leq 40$ in \autoref{fig:ConvergenceStokes2} for the sake of readability. The figure clearly shows that almost all of the PPOD modal energy is contained within the first $\sim 4 \%$ of modes. Moreover, it is observed that the rate of convergence towards unity increases as the Stokes number increases. 

There are several contributing factors to the observed behaviour of convergence. Firstly, the particles characterized by higher Stokes numbers are heavier, thus requiring more energy to be accelerated. Due to inertial filtering, the velocities of these particles fluctuate less around the mean (ensemble) velocity compared to lower Stokes number particles (\cite{ayyalasomayajula2008}; \cite{salazar2012}). Secondly, as seen in \autoref{fig:TKE} the increasing attenuation of TKE for increasing Stokes numbers implies a less energetic fluid surrounding the higher Stokes number particles, and the higher Stokes number particles are thus accelerated by smaller energies than the lower Stokes number particles. Thirdly, for lower Stokes numbers the small scale turbulent structures of the surrounding fluid are more energetic (\autoref{fig:spectrum}). The particles are in these cases accelerated by a wider range of turbulent structures resulting in more fluctuating particle velocities. Ultimately, these factors imply an increase in fluctuating particle velocities for low Stokes numbers compared to higher Stokes numbers, and hence a wider range of PPOD modes are required to account for these particle dynamics. The modal energy is thus more widely distributed for the lower Stokes number case, decreasing the convergence rate of $E_{a,modal}$.

\subsection{Component decomposition}
In stationary flows it is commonly accepted that fluid- and particle velocities may appropriately be decomposed with Fourier modes spanning the temporal domain (\cite{Tchen1947}; \cite{csanady1963}; \cite{hinze1975}; \cite{glauser1992}; \cite{delville1999}; \cite{citriniti2000}; \cite{johansson2002}; \cite{iqbal2007coherent}; \cite{muralidhar2019spatio}). The Fourier decomposition is applied such that each velocity component is decomposed separately. In analogy to this we now apply PPOD componentwise, i.e. with dimension $D=1$ we extract $M=DN_t =400$ modes and eigenvalues separately for the particle velocities in coordinate directions $x_1$, $x_2$ and $x_3$. \autoref{fig:Component}a shows up until $m = 40$ the convergence rate of $E_{a,modal}$ for component PPOD applied to the case $St_0 = 0.25$. Since the particles are suspended in decaying HIT, there is not a preferential direction, and the convergence rates are equivalent for all velocity component.

In \autoref{fig:Component}b the parallelity of the extracted modes is assessed by evaluating
\begin{align}
    P^{i,j} _{\alpha, \beta} = |(\boldsymbol{\varphi} ^i _\alpha,  \boldsymbol{\varphi} ^j _\beta)|, \quad i,j \in [1:3], \, \alpha,\beta \in [1:M], 
\end{align}
where $\boldsymbol{\varphi}^{i} _\alpha$ is the $\alpha$'th mode extracted for coordinate direction $x_i$. When $P^{i,j} _{\alpha, \beta} = 1$ the modes are completely parallel, whereas $P^{i,j} _{\alpha, \beta} = 0$ indicates orthogonality. The figure shows that along the diagonal ($\alpha = \beta$) there is almost complete parallelity for low mode numbers ($\alpha \leq 20$), signifying that the mode-sets extracted are basically the same. For higher mode numbers this is not the case, however as seen in \autoref{fig:Component}a these modes carry little energy, and they account for ensemble-specific variance rather than dominating particle dynamics. The importance of these modes is thus negligible, and it may be concluded that PPOD analysis of velocities in coordinate direction $x_i$ in decaying HIT yields the same qualitative results regardless of the value of $i$. Although only shown here for $St_0 = 0.25$, upon closer inspection of the data it is found that this conclusion may be drawn for every Stokes number considered, and similarly for fluid velocity modes extracted with component TPOD. For the remainder of this work, we will hence consider component PPOD and TPOD applied to velocities in coordinate direction $x_1$, and consider the results representative of all coordinate directions. 

\subsection{Modes}
\begin{figure*}
    \centering
    \includegraphics[scale=0.5]{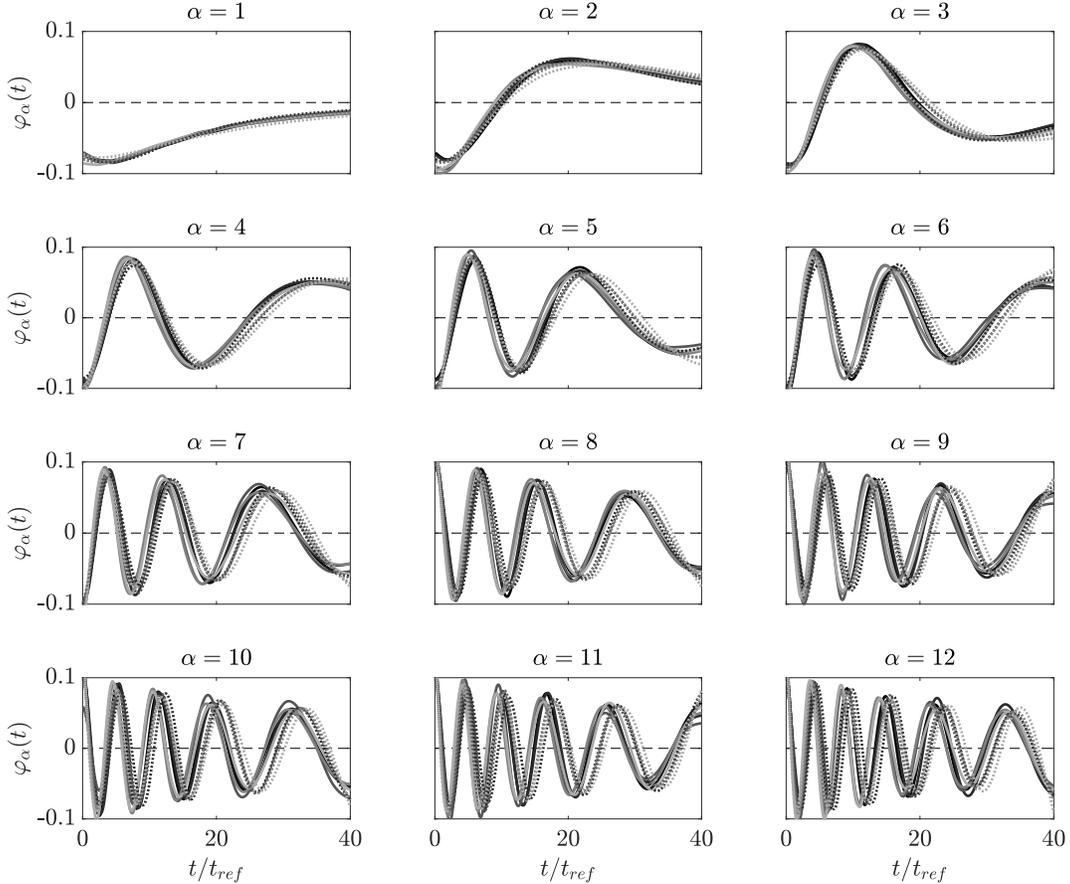}
    \caption{Modes ($\boldsymbol{\varphi}_\alpha$, $\alpha \in [1:12]$) extracted with TPOD (solid) and PPOD (dotted) for each simulation case. The nuance indicates Stokes number where the Stokes number increases from darker to lighter grey. The black solid lines are the TPOD modes for the single-phase case.}
    \label{fig:Modes}
\end{figure*}

A sample of the modes extracted with component TPOD (solid) for both the single- and multiphase simulations are shown in \autoref{fig:Modes} alongside a corresponding sample of the modes extracted with PPOD (dotted) in the multiphase cases. The modes are shown as functions of $t$, where $\boldsymbol{\varphi}_\alpha(t)$ denotes the element of $\boldsymbol{\varphi}_\alpha$ connected to sample time $t$. All mode-sets resemble slightly damped harmonic oscillators, where the local wavelength of each mode increases over time. The damping of amplitude may be attributed to the temporal decay of TKE (\autoref{fig:TKE}). The increase of wavelength shows that the energy-optimal modes for the decomposition of fluid- and particle velocities in decaying HIT are not Fourier modes, keeping in mind that in the stationary HIT case the componentwise TPOD and PPOD modes would be well approximated by Fourier modes (see e.g. \cite{aubry1991beauty} for TPOD modes).

\begin{figure}
    \centering
    \includegraphics[scale=0.45]{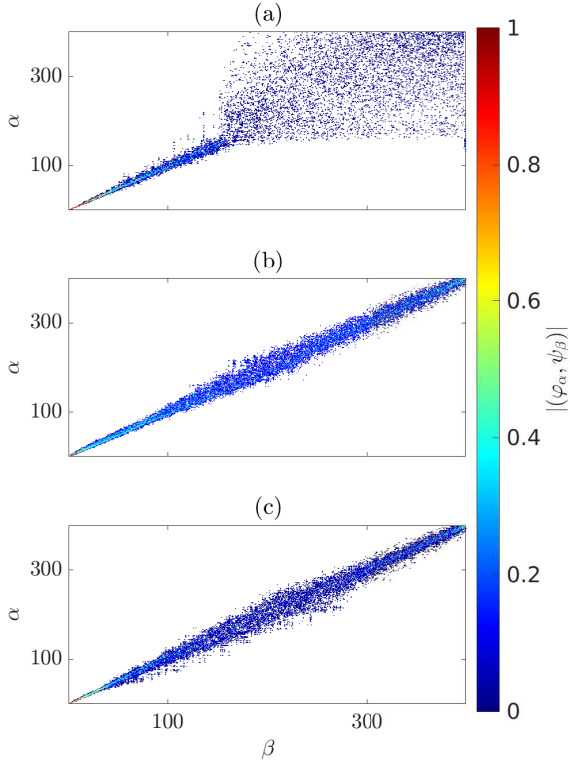}
    \caption{Parellelity between $\boldsymbol{\varphi}$ denoting the TPOD modes extracted for $St_0 = 0.25$ and $\boldsymbol{\psi}$ denoting the (a) TPOD modes of the single-phase simulation (b) TPOD modes extracted for $St_0 = 3.0$ and (c) PPOD modes extracted for $St_0 =0.25$. }
    \label{fig:ModeParallelity}
\end{figure}

A high correlation between all POD mode-sets is observed, indicating that the energetically dominating fluid- and particle dynamics do not vary considerably across Stokes numbers. This point is further investigated in \autoref{fig:ModeParallelity} where the parallelity between $\boldsymbol{\varphi}$ and $\boldsymbol{\psi}$ is evaluated. Here, $\boldsymbol{\varphi}$ is chosen as a reference mode-set given by the TPOD modes extracted from the $St_0 = 0.25$ simulation, and $\boldsymbol{\psi}$ represents the TPOD mode-set of the single-phase case (\autoref{fig:ModeParallelity}a), the TPOD mode-set of the $St_0=3.0$ case (\autoref{fig:ModeParallelity}b), and the PPOD mode-set of the $St_0 = 0.25$ case (\autoref{fig:ModeParallelity}c). Though figures \ref{fig:ModeParallelity}b and \ref{fig:ModeParallelity}c do not exhibit complete parallelity between $\boldsymbol{\varphi}$ and $\boldsymbol{\psi}$, they still illustrate a strong parallelity at lower mode numbers, and linear dependency of similarly numbered modes at higher mode numbers. This underlines that fluid- and particle dynamics are fairly similar across phase and Stokes number. \autoref{fig:ModeParallelity}a also exhibits strong parallelity at lower mode numbers, whereas $\boldsymbol{\varphi}$ is linearly dependent on many $\boldsymbol{\psi}$-modes for higher mode numbers. This shows that the dominating dynamics between the single- and multiphase flow are similar, and suggests that the two-way coupling information which is not captured in the single-phase modes is embedded in the higher numbered POD modes of the multiphase mode-sets. 

The high parallelity observed raises an important question: do the extracted bases span the same vector space? As briefly noted in \autoref{sec:POD} the extracted POD-bases are not necessarily complete in $\mathcal{H}$, and it is therefore not guaranteed that the velocities of one ensemble may be fully decomposed by the modes extracted from another ensemble. However, if one mode-set $\boldsymbol{\varphi}$ can be completely decomposed by another mode-set $\boldsymbol{\psi}$, then the ensemble members generating $\boldsymbol{\varphi}$ can also be fully decomposed by $\boldsymbol{\psi}$. In total nine mode-sets are extracted in the current work (five TPOD and four PPOD) and it turns out that each of these can fully reconstruct (down to machine precision) the other eight mode-sets, i.e. 
\begin{align}
   \left| \left| \boldsymbol{\varphi}_\beta - \sum \limits_{\alpha=1}^{M} (\boldsymbol{\varphi}_\beta, \boldsymbol{\psi}_\alpha) \boldsymbol{\psi}_\alpha \right| \right| = 0, \quad \beta \in [1:M].
\end{align}
All mode-sets therefore span the same vector space, and both particle- and fluid velocities may be expanded in the same basis. Though not shown here, $\boldsymbol{u}_{f@p}$ (equation \eqref{eq:ParticleMotion}) may also be fully decomposed with respect to the POD bases extracted in the current study. This enables the use of a single empirically determined basis to be used for the expansion of all data sets in question, illustrating the versatility of the PPOD method in the application to multiphase flows.

\subsection{Response - velocity}
Following the procedure of \cite{csanady1963} it may be shown for statistically stationary flows that the Fourier Lagrangian spectrum of the particle velocity, $E_p$, is connected to the equivalent spectrum of the fluid velocity at particle position, $E_{f@p}$, by a response function $H^2$. The relation is given by
\begin{align}
    E_p(\alpha) = H^2 (\alpha) E_{f@p} (\alpha), \quad \alpha \in [1:M], \label{eq:ResponseEnergy}
\end{align}
where $E_p(\alpha)$ and $E_{f@p}(\alpha)$ is the ensemble averaged energy connected to the $\alpha$'th Fourier mode for respectively the dispersed- and carrier phase. An analytic expression for $H^2$ may be found by replacing $\boldsymbol{v}_p$ and $\boldsymbol{u}_{f@p}$ in equation \eqref{eq:ParticleMotion} with their Fourier expansions. This was also done by \cite{berk2021} where they showed that
\begin{align}
    H^2(\alpha) = \frac{1}{1+(\omega(\alpha) \tau _p )^2}\,, \quad \alpha \in [1:M]\,. \label{eq:Response}
\end{align}
Here $\omega (\alpha)$ is the angular frequency of the $\alpha$'th Fourier mode and 
\begin{align}
    \tau _p = \frac{\rho _p d_p ^2}{18 \mu _f} \left( 1 + 0.15 Re_p^{0.687} \right) ^{-1}. \label{eq:ParticleResponseTime}
\end{align}
To the authors knowledge, no analytic expressions exist for the response function $H^2$ in non-stationary flows at the time of writing. However, we may study the fraction $E_p(\alpha) / E_{f@p}(\alpha)$ and let this serve as an empirical response function. 

Considering Fourier modes as a special case of POD modes, i.e. those derived empirically for a statistically stationary flow (\cite{glauser1992}), we conjecture that some of the properties of the Fourier basis may also apply to the POD basis in general. Indeed, for select test functions representing stationary dynamics, \cite{hodzic2022} observed a high correlation between the POD eigenspectrum and the analytical Fourier spectrum. Interestingly, the correlation exceeded that of the analytical Fourier spectrum and the spectrum of the discrete Fourier transform (DFT), indicating a close spectral symmetry between the analytical Fourier basis and the POD basis in \textit{locally} statistically stationary flows. 

\begin{figure}
    \centering
    \includegraphics[scale=0.44]{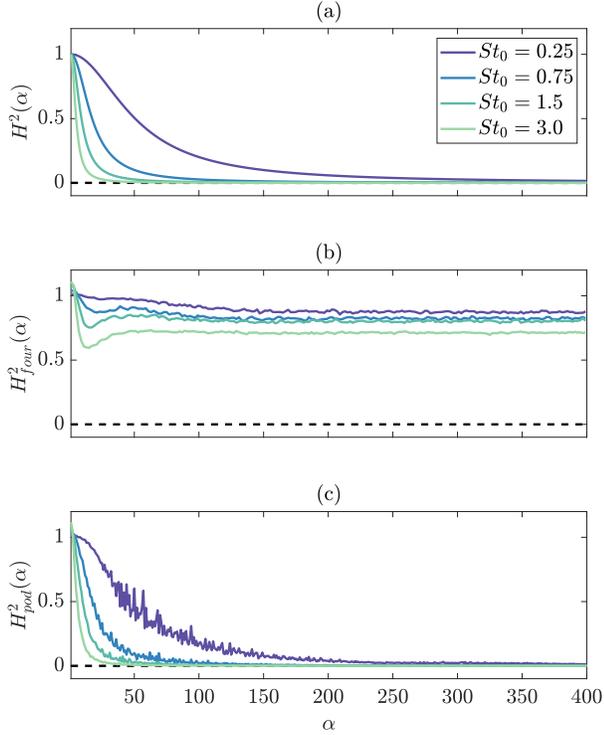}
    \caption{Analytic response function derived for stationary flows is shown in (a) whereas (b) and (c) are respectively the Fourier- and POD empirical response functions for the current non-stationary flow.}
    \label{fig:Response}
\end{figure}

Based on these considerations, and recalling that PPOD modes may, in our simulations, completely expand both $\boldsymbol{v}_p$ and $\boldsymbol{u}_{f@p}$, we hypothesize that the empirical response function based on PPOD modes follow a trend similar to $H^2$ (equation \eqref{eq:Response}). The hypothesis is validated in \autoref{fig:Response} where $H^2$, $H_{four}^2$ and $H_{pod}^2$ are shown. Here $H_{four}^2$ (\autoref{fig:Response}b) is shown as a reference case, representing the empirical response function where $E_p$ and $E_{f@p}$ are computed based on Fourier modes. $H_{pod}^2$ (\autoref{fig:Response}c) represents the empirical response function computed based on the PPOD modes of each simulation. For $H^2$, $\tau _p = \tau _p (t_0)$ is used in each case, although the quantity is dependent on time since the studied flow is non-stationary. However, for the Stokes numbers and temporal domain considered it is observed that
\begin{align}
    \frac{|\tau _p (t_0) - \tau _p (t_{N_t-1})|}{\tau _p(t_0)} \leq 8\%\,.
\end{align}
It is therefore assumed that the choice of $\tau _p$ is reasonably representative of the dynamics over the entire temporal domain in each simulation case.

Though equation \eqref{eq:Response} is derived based on the Fourier transform, Figures \ref{fig:Response}a and \ref{fig:Response}b show little correlation between $H^2$ and $H_{four}^2$. The inherent periodicity of Fourier modes justifies this result, since expansion of non-periodic signals will lead to spectral leakage. We are studying decaying HIT, and as a consequence, the fluid- and particle velocity signals are not periodic and the Fourier modes do not form an appropriate basis for the expansion of these signals (\cite{lumley2007}).

Conversely, Figures \ref{fig:Response}a and \ref{fig:Response}c show a high correlation between $H^2$ and $H_{pod}^2$, as hypothesized. Feasibly, the result reflects some deeper spectral symmetry related to energy optimality, from which equation \eqref{eq:Response} follows, rather than the equation strictly following from the properties of Fourier modes. 

\begin{figure}
    \centering
    \includegraphics[scale=0.41]{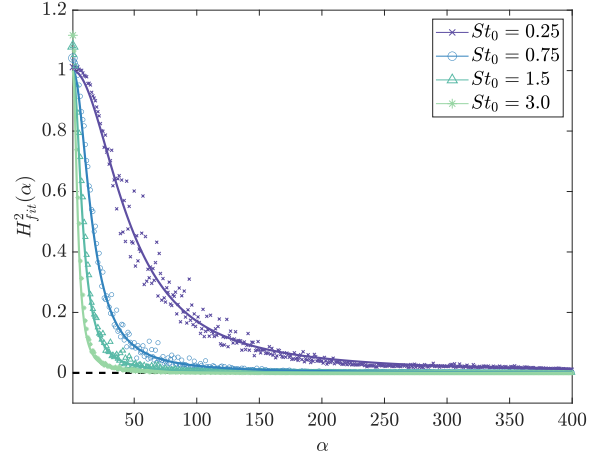}
    \caption{Fit of analytic response function (solid) to empirical POD response function (markers).}
    \label{fig:FittedResponse}
\end{figure}

In \autoref{fig:FittedResponse} the correlation is more clearly illustrated by the depiction of $H_{pod} ^2$ (markers) and $H^2_{fit}$ (solid). Here $H_{fit}^2$ is a least squares fit of $H^2$ to $H^2_{pod}$ given by
\begin{align}
    H_{fit}^2 (\alpha) = \frac{1}{1+((\alpha - 1)\omega^* \tau _p )^2}\,, \quad \alpha \in [1:M]\,. \label{eq:ResponseFunction}
\end{align}
The product $(\alpha - 1) \omega ^*$ represents a "POD-frequency", and the fitting parameter $\omega^* = 9.0647$ is found through minimization of the objective
\begin{align}
    \min \limits_{\omega^*} \, \sum \limits_{\tau _p} \sum_{\alpha} \left|\left| \frac{H_{fit}^2(\alpha) - H^2_{pod}(\alpha)}{H_{fit}^2(\alpha)} \right|\right|_2\,. \label{eq:objective}
\end{align}
Summation over $\tau _p$ represents the fitting of $\omega^*$ to the data of all Stokes numbers simultaneously. The figure shows a clear connection between the PPOD empirical response function and a modified version of the analytical model \eqref{eq:Response}. These results show potential for the ability to approximate the modal energy of the carrier phase sampled at the particle position in decaying HIT, directly from the Stokes number and particle velocities.

%

\subsection{Response - relative velocity}
\begin{figure}
    \centering
    \includegraphics[scale=0.41]{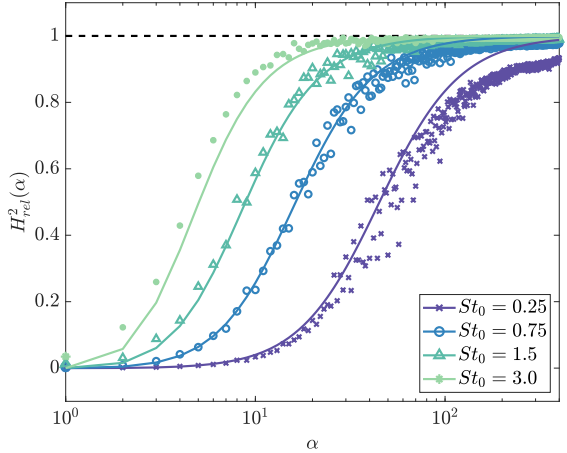}
    \caption{$H_{rel}^2$ (solid) plotted against the fraction $E_{rel} / E_{f@p}$ (markers) shows a fairly good fit for Stokes numbers $St_0 \leq 1.5$ and $\alpha \leq 30$.}
    \label{fig:RelativeResponse}
\end{figure}

\cite{csanady1963} derived a relation between the mean square relative velocity and $E_{f@p}$. Inspired by this, and extending the considerations of the previous section, we conjecture that 
\begin{align}
    E_{rel} ( \alpha ) = H_{rel}^2 ( \alpha ) E_{f@p} ( \alpha )\,, \quad \alpha \in [1:M]\,,  \label{eq:RelativeEnergy}
\end{align}
where $E_{rel}(\alpha)$ is the ensemble averaged energy of the relative velocity, $\boldsymbol{u}_{rel} = \boldsymbol{u}_{f@p} - \boldsymbol{v_p}$, connected to the $\alpha$'th POD mode (PPOD), and
\begin{align}
    H_{rel} ^2(\alpha) = ((\alpha - 1) \omega ^{*} \tau _p )^2 H_{fit}^2 ( \alpha)\,, \quad \alpha \in [1:M]\,. \label{eq:RelativeResponse}
\end{align}
\autoref{fig:RelativeResponse} depicts the empirically evaluated fraction $E_{rel}(\alpha) / E_{f@p}(\alpha)$ (markers) and the fit $H_{rel}^2 (\alpha)$ (solid). The quantities are plotted against the log-scaled $\alpha$ to highlight the fit at the energetically dominant modes. Visual inspection shows a good model fit for $\alpha \leq 30$ and $St_0 \leq 1.5$, whereas the fit and the empirical results diverge at higher mode numbers. Noting that the modes $\boldsymbol{\varphi}_\alpha$, $\alpha \in [1:30]$, account for more than 99\% of the total modal energy in each case, it can be argued that the fit is appropriate at the energetically dominant modes for $St_0 \leq 1.5$. For $St_0 = 3.0$ the model fit at the energetically dominant modes is not on point. This suggests a range of Stokes numbers at which the model is appropriate. However, in all cases, the trend of $E_{rel} / E_{f@p}$ is similar to that of $H_{rel}^2$, highlighting the similarities between the models derived through Fourier analysis of stationary flows and POD analysis of the current non-stationary flow. 

Equations \eqref{eq:ResponseEnergy}, \eqref{eq:ResponseFunction}, \eqref{eq:RelativeEnergy} and \eqref{eq:RelativeResponse} provide a method for approximating $E_{f@p}$ and $E_{rel}$ in particle-laden decaying HIT based on particle velocity measurements. The method requires knowledge of $\tau _p$, and if this quantity changes significantly over the considered temporal domain, or if it has rapid fluctuations, the current accurateness of the model may deteriorate since it was assumed that $\tau _p = \tau _p (t_0)$ for the generation of the fit.

%
%

\section{Conclusions} \label{sec:Conclusion}
A study of the temporal dynamics of two-way coupled particle-laden decaying HIT for various Stokes numbers was conducted. Using time-focalized formulations of POD -- TPOD and PPOD for respectively the decomposition of fluid- and particle velocities -- sets of energy-optimal modes were extracted representing the temporal dynamics of the two phases. For both phases it was observed that the extracted modes resembled damped harmonic oscillators, where the local wavelength of each mode increased over time. Moreover, the modes exhibited a high correlation in the dominating dynamics between the carrier- and dispersed phase. 

The TPOD eigenspectrum of each simulation was inspected and compared to the eigenspectrum of a corresponding single-phase simulation. A distinct difference of shape between the single- and multiphase spectra was observed. In addition, the TPOD spectra were compared to the Fourier turbulence energy spectrum generated at the final time step of each simulation. Here an increase of energy at high wavenumbers of the turbulence spectrum was observed to correlate with a relative increase in the TPOD eigenspectrum at high mode numbers.

It was demonstrated that the POD mode-set extracted from the velocity of one phase could span the velocity of both phases. Therefore, the Lagrangian spectrum based on PPOD modes could be computed for both the carrier- and dispersed phase. A relation between these spectra was evaluated empirically giving rise to analytical expressions of response functions in a PPOD frame of reference. The response functions related the modal energy of the inertial particles to that of the surrounding fluid through simple expressions dependent on the Stokes number. Notably, the expressions, fitting the data of the current non-stationary flow, resembled those derived through Fourier analysis of stationary flows. This suggested a deeper symmetry between POD and Fourier spectra. 

The current PPOD analysis was applied to an ideal test case of a non-stationary flow. The results outlined, and the theoretical applicablity of PPOD to any non-stationary flow, indicate that PPOD analysis may provide insightful dynamical information for the Lagrangian dynamics of alternative flows in future studies of particle-laden turbulence.

\begin{acknowledgments}
AH and CMV acknowledge financial support from the European Research council: This project has received funding from the European Research Council (ERC) under the European Unions Horizon 2020 research and innovation program (grant agreement No 803419). 

MS acknowledges financial support from the Poul Due Jensen Foundation: Financial support from the Poul Due Jensen Foundation (Grundfos Foundation) for this research is gratefully acknowledged.
\end{acknowledgments}

\section*{Declaration of interest}
The authors report no conflicts of interest. 

\section*{Data Availability Statement}
The data that forms the basis of this study is available from the corresponding author upon reasonable request. 

\nocite{*}
\section*{References}
\bibliography{zReferences}

\end{document}